\documentclass[aps,prd,twocolumn,showpacs,floatfix,
  superscriptaddress,amsmath]{revtex4}
\usepackage{epsf}              %
\usepackage{amscd}              %
\usepackage{amsmath}            
\usepackage{graphicx} %
\usepackage{comment}            %
 
\usepackage{amssymb,hyperref, amsmath}
\usepackage[dvips]{color}

\input xy
\xyoption{all}

\newcommand{\beq}{\begin{equation}}
\newcommand{\eeq}{\end{equation}}
\newcommand{\bea}{\begin{eqnarray}}
\newcommand{\eea}{\end{eqnarray}}

\newcommand{\benn}{\begin{displaymath}}
\newcommand{\eenn}{\end{displaymath}}

\newcommand{\tw}{\textwidth}

\newcommand{\ig}{\includegraphics}
\def\slashchar#1{\ensuremath{                   %
   \setbox0=\hbox{${}#1{}$}                     
   \dimen0=\wd0                                 
   \setbox1=\hbox{/} \dimen1=\wd1               
   \ifdim\dimen0>\dimen1                        
   \rlap{\hbox to \dimen0{\hfil/\hfil}}         
   {}#1{}                                       
   \else                                        
   \rlap{\hbox to \dimen1{\hfil${}#1{}$\hfil}}  
   /                                            
   \fi}}                                        %
\def\simge{
 \mathrel{\rlap{\raise 0.511ex
  \hbox{$>$}}{\lower 0.511ex \hbox{$\sim$}}}}
\def\simle{
 \mathrel{\rlap{\raise 0.511ex
  \hbox{$<$}}{\lower 0.511ex \hbox{$\sim$}}}}

\begin{document}

\hbox{UMD-DOE-40762-472,~ ~JLAB-THY-10-1170, ~~DESY-10-057 \\}
\vspace{0.1in}

\title{Nucleon, $\Delta$ and $\Omega$ excited state spectra in $N_f$=2+1 lattice QCD}

\author{J.~ Bulava}
\affiliation{NIC, DESY, Platanenallee 6, D-15738, Zeuthen, Germany}
\author{R.~G.~Edwards} 
\affiliation{Thomas Jefferson National Accelerator Facility,
Newport News, VA 23606, USA}
\author{E.~Engelson}  
\affiliation{Department of Physics, University of Maryland, College Park, MD 20742, USA}
\author{B.\ Jo\'{o} }
\affiliation{Thomas Jefferson National Accelerator Facility,
Newport News, VA 23606, USA}
\author{H-W. Lin}
\affiliation{Department of Physics, University of Washington, Seattle, WA 98195-1560}
\author{C.~ Morningstar}
\affiliation{Department of Physics, Carnegie Mellon University,
Pittsburgh, PA 15213, USA}
\author{D.~G.~Richards}
\affiliation{Thomas Jefferson National Accelerator Facility,
Newport News, VA 23606, USA}
\author{S.~ J.~ Wallace}
\affiliation{Department of Physics, University of Maryland, College Park, MD 20742, USA}
\collaboration{ for the Hadron Spectrum Collaboration }
\noaffiliation 
\begin{abstract}
The energies of the excited states of the Nucleon, $\Delta$
and $\Omega$ are computed in lattice QCD, using two light quarks and one strange quark
on anisotropic lattices.
The calculation is performed at three values of the light quark mass, 
corresponding to pion masses $m_{\pi}$ = 392(4), 438(3) and 521(3) MeV. 
 We employ the variational method with a large basis of interpolating operators enabling
 six energies in each irreducible representation of the lattice to be distinguished clearly.
  We compare our calculation with the low-lying experimental spectrum, with which we find
  reasonable agreement in the pattern of states.  The need to include operators that
  couple to the expected multi-hadron states in the spectrum is clearly identified.
    
\end{abstract}
 \maketitle

\section{Introduction}

The theoretical determination of the spectrum of baryon resonances
from the fundamental quark and gluon degrees of freedom is an
important goal for lattice QCD.  Recently a number of groups have performed 
lattice computations for $N_f = 2+1$ QCD with two light-quark flavors (up and down)
and one more massive quark flavor (strange).\cite{Lin:2009, Durr:2008, Allton:2008, Yamazaki:2009, Aubin:2005, Alexandrou:2009, Gattringer:2009, Bazavov:2010} Different actions have been used
and several analyses achieve a pion mass close to the physical limit.   A recent analysis
used a reweighting technique to perform calculations at the physical pion mass.\cite{Aoki:2010} 
One focus of attention has been the determination of the lowest baryon mass 
for each isospin and strangeness.  Good agreement 
has been achieved between different calculations.  The experimental
masses are reproduced typically with discrepancies ranging from 1\% to 8\%.   

Another focus of attention is the excited state spectrum of baryons. 
Recent works have progressed beyond quenched QCD~\cite{Basak:2007,Burch:2006cc} which omits the effects of 
light quarks from the gauge ensembles. Two light-quark flavors ($N_f = 2$) 
were used in Ref.~\cite{Bulava:2009}.  In this work  we use ensembles of gauge configurations
developed in Ref.~\cite{Lin:2009} for $N_f=2+1$ QCD with dynamical light and strange quarks.

There are ongoing experimental programs aimed at determining the spectra and properties of excited 
baryons at the Thomas Jefferson National Accelerator Facility.  Reference~\cite{Burkert:2009} provides
an overview of some recent experimental results.
 
Excited baryon states can be quite massive and a small lattice spacing in
the time direction is best for observing their signals, which decrease exponentially with time.
 If the same lattice spacing were used in spatial and time directions the computational cost would be 
 unnecessarily high. The Hadron Spectrum Collaboration has undertaken a program to solve QCD using
anisotropic lattices that have a smaller spacing, $a_t$, in the time direction. In this work we use
16$^3\times$128 lattices with a renormalized anisotropy $\xi$ = 3.5~\cite{Edwards:2008}, i.e.,  
$a_t$ = $a_s/$3.5, where $a_s$ is the spatial lattice spacing.  A Symanzik-improved gauge and a clover-improved Wilson fermion action have been used to generate gauge ensembles for $N_f=2+1$ QCD.~\cite{Lin:2009}   The three ensembles used in this work have pion masses of 392(4) MeV, 
438(3) MeV and 521(3) MeV.  
 The effects of charm, bottom and top quarks are neglected because the baryons under study
have only light quarks in their valence structures and the loop contributions of the neglected quarks are suppressed by their large masses.  

For families of particles with given isospin and strangeness, spectra are calculated 
in the six double-valued irreducible representations (irreps) of 
the octahedral group.  There are three irreps for
even-parity that are labeled with a $g$ subscript ({\it gerade})
and three for odd-parity that are labeled with a $u$ subscript
({\it ungerade}). They are: $G_{1g}, H_g, G_{2g}, G_{1u}, H_u$ and
$G_{2u}$.  Operators that transform according to one of these
irreps do not mix with those of other irreps because of the octahedral symmetry. 

  A large basis of interpolating field operators is needed in order to 
  extract the spectrum of excited states.  We have developed many such operators
  for each isospin, strangeness and octahedral irrep in previous works.\cite{Basak:2005}
  In this work we select sets of 7 to 11 operators in each irrep.   
Continuum values of total angular momenta show up in lattice
simulations as patterns of degenerate energies in the continuum
limit that match the patterns in Table~\ref{table:subduction} for
the subduction of spin $J$ to the double-valued irreps of the
octahedral group.
\begin{table}
\caption{The number of occurrences of double-valued irrep
$\Lambda$ of the octahedral group for half-integer values of
continuum spin $J$. N is the dimension of the irrep. } \label{table:subduction}
\hspace{-0.2in}
\begin{tabular}{|ccccccccc|}
\hline
    &   & ~J~ &  ~$1/2$~&~$3/2$~&~ $5/2$ ~&~ $7/2$~&~ $9/2$~&~ $11/2$   \\
$\Lambda$  & N  & & ~&~~&~~&~ ~&~ ~&~   \\
\hline
$G_{1}$ ~&~2~& & 1 ~&~ 0 ~&~ 0 ~&~ 1 ~&~ 1 ~&~ 1    \\
$H $    ~&~4~& & 0 ~&~ 1 ~&~ 1 ~&~ 1 ~&~ 2 ~&~ 2  \\
$G_{2}$~& ~2~& & 0 ~&~  0 ~&~ 1 ~&~ 1 ~&~ 0 ~&~ 1  \\
\hline
\end{tabular}
\end{table}
For example, a state in one of the $G_2$ irreps is a signal for
the subduction of continuum spin $\frac{5}{2}$ or higher. Because spin
$\frac{5}{2}$ has six magnetic substates and the 
lattice $G_{2}$ irrep has dimension two, there must be four degenerate partner states 
from the four-dimensional $H$ irrep in order to realize the six linearly 
independent states that are required.  For spin $\frac{7}{2}$, there 
must be degenerate partner states in the $G_1$, $H$ and
$G_2$ irreps in order to realize the total of eight magnetic substates.

This paper is organized as follows. In
section~\ref{sec:lattices-etc.}, we review the lattices used and
the basic features of the action and quark masses, the baryon operators, 
the smearing of quark fields
based on eigenvectors of the gauge-covariant lattice Laplacian,
the pruning of operators to obtain suitable sets for calculations of
matrices of correlation functions and the variational method with eigenvectors fixed at one time value.   As we have shown in previous work~\cite{Bulava:2009}, smearing 
of the source and sink operators is important for calculations of baryon spectra because it 
reduces contributions from short wavelength fluctuations.  We use the recently developed distillation method of Ref.~\cite{Peardon:2009} in which
source and sink quark operators are smeared by applying eigenvectors of the gauge-covariant 
Laplacian,  $-\nabla^2$.  

The orthogonality of the eigenvectors produces a factorization of the problem.  One part consists of the
smeared baryon operators based on $N_{\rm eig}$ eigenvectors of the scalar Laplacian $-\nabla^2$ on each time slice for each
quark.  The other consists of correlation functions corresponding to the parallel transport of
 smeared quark operators formed from the eigenvectors of the Laplacian from the source time slice
 to the sink time slice.   
These elementary quark correlation functions are called ``perambulators'' for short. 
They are universal in the sense that they may be used for any baryon or 
meson operators. The distillation method provides an `all-to-all' 
calculation in the sense that propagators are calculated from all $N_{\rm eig}$ eigenvectors 
of $-\nabla^2$ at the source to all $N_{\rm eig}$  eigenvectors at the sink. 

In Section~\ref{sec:N-spectrum-392} 
we report calculations for the nucleon and show details of the fits to obtain energies and
the results for the spectrum at the lowest pion mass, 392 MeV.
In Section~\ref{sec:N-D-Om-spectra} we report summary results for the nucleon, $\Delta$ and
$\Omega$ spectra for all three pion masses and comparisons with experimental resonances.  The $\Omega$ spectrum of low-lying excited states
is interesting because little experimental information is available.  
It also helps to set the overall scale for baryon spectra, for which we use the ground state $\Omega$ baryon mass at each value of $m_{\pi}$.

 Section~\ref{sec:summary} gives a summary of the results. 

\section{Lattices, operators and matrices of correlation functions}
\label{sec:lattices-etc.}

    References~\cite{Lin:2009} and~\cite{Edwards:2008} presented the tuning of quark masses and other parameters for $N_f$=2+1 QCD
    on the anisotropic $16^3\times 128$ lattice used in this work.  Reference~\cite{Peardon:2009}
    presented the method of smearing hadronic operators using an expansion 
    in terms of eigenvectors of the three-dimensional lattice Laplacian.  Because those references 
provide the foundation for the present work, we summarize their
    findings in this section.  
    
    Table~\ref{tab:quarkmasses} shows the three sets of quark masses from Ref.~\cite{Lin:2009} used in this work together with values for the pion mass, the kaon mass and 
    the $\Omega$ baryon mass.  The value of the pion mass in MeV units is obtained by using the
    $\Omega$ mass at each point to set the scale.
We also show the number of gauge configurations used.  For each configuration we compute
correlation functions from four different time sources and average them in order to take account of correlations.  

     \begin{table}
\caption{ Lattice parameters.  Three sets of values (in temporal lattice units) of the light 
quark mass, $m_{\ell}$, the strange quark mass, $m_s$ and the resulting $\pi$ meson, $K$ meson and 
$\Omega$-baryon masses.  The corresponding values of 
$m_{\pi}$ in MeV are given in the last row based on using Eq.~(\ref{eq:scale}).
  }
   \begin{center}
    \begin{tabular}{|c|ccc|}
   \hline
   ensemble  &  1  &  2  &  3  \\
   \hline
   $m_{\ell}$ & $-$.0840 & $-$.0830 & $-$.0808 \\
   $m_{s}$  & $-$.0743 & $-$.0743 & $-$.0743 \\
   Volume    & 16$^3\times$ 128 & 16$^3\times$ 128 &16$^3\times$ 128 \\
     $N_{\rm cfgs}$  &   344   &   570  &  481   \\ 
     $t_{\rm sources}$ & 4  &  4  & 4    \\    
\hline
   $m_{\pi}$ & 0.0691(6) & 0.0797(6) & 0.0996(6) \\
   $m_K$    & 0.0970(5)  & 0.1032(5) & 0.1149(6) \\
   $m_{\Omega}$ & 0.2951(22) & 0.3040(8) & 0.3200(7) \\
\hline
   $m_{\pi}$ (MeV) &  392(4) &   438(3) &  521(3) \\
   \hline
\end{tabular}\label{tab:quarkmasses}
 \end{center}   \end{table}
     
        Baryon operators used in this work were developed in 
Ref.~\cite{Basak:2005}.  For the baryons that we have considered, the 
single-site (SS) forms of the operators are given in Table~\ref{tab:baryon_SSops}.  
Most operators incorporate gauge-covariant displacements of the quarks relative to one another
 in order to obtain nontrivial shapes.  The displaced operators are projected to irreps of the 
        octahedral group by summing over all lattice rotations applied to the shapes 
and spins with coefficients that project out the irrep operators.  Many 
operators are so obtained and in the final step they are ``pruned'' to sets of 
between 7 and 11 operators for each $N$ and $\Delta$ irrep.  For $\Omega$, we use the same operators as for 
$\Delta$ except that the number is limited to 6 in each irrep.  The operators 
are selected to have good signal to noise characteristics and correlation matrices with low 
condition numbers at a time close to the source.  Low condition number 
(ratio of largest to smallest eigenvalue) 
is a proxy for linear independence, which follows because two linearly dependent 
operators would give a zero eigenvalue for a matrix of correlation functions and 
an infinite condition number.    A full listing of operators is available upon request.
     
\renewcommand{\a}{\alpha}
\renewcommand{\b}{\beta}
\renewcommand{\c}{\gamma}
\begin{table}[h]
\begin{center}
\caption{Baryons and the corresponding three-quark
elemental operators. Columns 1 to 4 show the symbol, isospin,
strangeness, and the form of elemental single-site
operators used.   The last three columns show the numbers of embeddings of single-site operators
with irreps $G_{1g}, G_{2g}$ and $H_{g}$, respectively.  \label{tab:baryon_SSops} }
\begin{tabular}{|c|c|c|c|c|c|c|}
\hline
     & $I$ & $S$ & $B_{\a\b\c}$& $G_{1g}$ & $G_{2g}$ & $H_{g}$  \\
\hline 
$N$ &      $1/2$ & $0$ & $(u_\a d_\b - d_\a u_\b) u_\c/\sqrt{2}$ &  3  &  0   &  1  \\
$\Delta$&  $3/2$ & $0$ & $( u_\a  u_\b  d_\c + u_\a  d_\b  u_\c +  d_\a  u_\b  u_\c)/\sqrt{3}$  & 1  &  0   &  2   \\
$\Omega$  & $0$  & $-3$ & $( s_\a  s_\b  s_\c + s_\a  s_\b  s_\c +  s_\a  s_\b  u_\c)/\sqrt{3}$ & 1  &  0   & 2    \\ 
\hline
\end{tabular}
\end{center}
\end{table}

       Table~\ref{tab:pruned_ops}  indicates the number and type of pruned operators for each irrep for 
        $N$,  $\Delta$ and $\Omega$ baryons 
        used in this work following the conventions of Ref.~\cite{Basak:2005}.  The types of
        operators are: SS (single-site); SD (singly displaced) with one quark gauge-covariantly displaced 
        from the other two; DDI (doubly-displaced-I) with two quarks displaced in opposite directions from the third; and TDT (triply-displaced-T) with all quarks displaced to create a T shape.  
\begin{table}[h]
\begin{center}
\caption{Numbers of operators of each type used in this work for N, $\Delta$ and $\Omega$ matrices are listed
for each irrep.  SS denotes single-site (local) operators, SD denotes singly-displaced operators, DDI denotes doubly-displaced-I operators, DDL denotes doubly-displaced-L operators and TDT denotes triply-displaced-T operators.  Gauge-covariant displacements are used. \label{tab:pruned_ops} }
\begin{tabular}{|c|c|cccccc|}
\hline
   Baryon & Operator & $G_{1g}$ & $G_{1u}$ & $H_g$& $H_u$ & $G_{2g}$ & $G_{2u}$  \\
                 &   type       &                    &                   &              &              &                   &                   \\
\hline 
$N$ &   SS    &   2    &   1    &  1    &   1   &  0   &   0 \\
$N$ &  SD     &   1    &   1    &  2    &   1   &  1   &   2 \\
$N$ &  DDI    &   0    &   2    &  1    &   3   &  3   &   2 \\
$N$ &  DDL  &    2    &   2    &  4    &   3   &  3   &   2 \\
$N$ &  TDT   &    2    &   4    &  3    &   1   &  4   &   2 \\
\hline 
$N$ &  total   &    7   &  10    & 11  &  9    &  11 &  8 \\
\hline
$\Delta$ &  SS     &   0    &   0    &   1    &   2   &  0   &   0 \\
$\Delta$ &  SD     &   3    &   3    &   2    &   0   &  4   &   2 \\
$\Delta$ &  DDI    &   2    &   2    &   1    &   2   &  0   &   0 \\
$\Delta$ &  DDL  &    3    &   3    &   2    &   2   &  4   &   4 \\
$\Delta$ &  TDT   &    2    &   1    &   3    &   4   &  3   &   3 \\
\hline
$\Delta$ &  total   &  10   &    9    &  9     &  10   &  11 &  9 \\
\hline
$\Omega$ &  SS     &   0    &   0    &   1    &   2   &  0   &   0 \\
$\Omega$ &  SD     &   3    &   3    &   2    &   0   &  4   &   2 \\
$\Omega$ &  DDI    &   2    &   2    &   1    &   2   &  0   &   0 \\
$\Omega$ &  DDL  &    1    &   1    &   2    &   2   &  2   &   4 \\
\hline
$\Omega$ &  total   &  6  &   6    &  6     &  6   &  6 &  6 \\
\hline
\end{tabular}
\end{center}
\end{table}

In order to reduce couplings to short-wavelength lattice fluctuations,   
smearing of the operators is performed as in Ref.~\cite{Peardon:2009}. 
The method is called ``distillation'' and it uses the eigenvectors of the 
gauge-covariant, three-dimensional Laplacian operator.  The k$^{th}$ eigenvector 
depends on a color index, a, and the spatial coordinates, ${\bf x}$ and is written as 
$v^{(k)}_{a{\bf x}}$.  It obeys the eigenvalue equation
\begin{equation}
\big(-\nabla^2\big)_{\bf x y}^{ab} v^{(k)}_{b,\bf y} = \lambda_k v^{(k)}_{a{\bf x}},
\end{equation}
 where $a$ and $b$ are color labels.  Ordering is imposed such that increasing $k$ 
corresponds to increasing eigenvalues $\lambda_k$.
 
 A sum over all $M = N_c \times N_x \times N_y \times N_z$ eigenvectors for a given lattice provides a decomposition of unity on time slice t, i.e.,
 \begin{equation}
 \sum_{k=1}^M v_{a,\bf x}^{(k)}(t)v_{b,\bf y}^{(k)\dag}(t) = \delta^{ab} \delta_{\bf x y}.
 \end{equation}
 A sum over only the $N_{\rm eig}$ lowest eigenvalues provides the distillation operator on time-slice t as 
 the following projection,
 \begin{equation}
  \Box^{ab}_{\bf xy}(t) = \sum_{k=1}^{N_{\rm eig}} v_{a,\bf x}^{(k)}(t)v_{b,\bf y}^{(k)\dag}(t).
  \end{equation}
  
Applying the projection to a quark field yields
\begin{eqnarray}
 \Box^{ab}_{\bf xy}(t) q_{\alpha}^b ({\bf y},t) 
  = \sum_{k=1}^{N_{\rm eig}} v_{a,\bf x}^{(k)}(t) \widetilde{q}_{\alpha}^{(k)}(t),
  \label{eq:project}
\end{eqnarray}
where the smeared field operator is defined by
\begin{equation}
\widetilde{q}_{\alpha}^{(k)}(t) =  v_{b,\bf y}^{(k)\dag}(t) q_{\alpha}^b ({\bf y},t) .
\end{equation}
 The smeared field operator involves a sum over repeated indices $b$ for color and {\bf y} for space. It 
 depends only on the Dirac index, $\alpha$, the Laplacian eigenvector label, k, and time, t.  It is of 
 rank $N_{\rm eig}\times N_{\sigma}$ vector on each time slice, where $N_{\sigma}$ is the number of
 spinor components, i.e., four.  A similar projection is used for
 displaced operators as discussed in Ref.~\cite{Peardon:2009}.
 
 Matrices of correlation functions are calculated as follows,
  \begin{equation}
 C_{ij}(t,t') = \sum_{\bf x y} \big\langle B_i({\bf x},t) B_j^{\dag} ({\bf y},t')  \big\rangle ,
 \end{equation}
 where,  using single-site operators,  
 \begin{equation}
B_i({\bf x},t) =  C_i^{\alpha\beta\gamma} \epsilon^{abc} q_{\alpha}^{a f_1} ({\bf x},t)
q_{\beta}^{b f_2} ({\bf x},t) q_{\gamma}^{c f_3}({\bf x},t).
\end{equation}
Superscripts $a$, $b$ and $c$ are color indices while $f_1$, $f_2$ and $f_3$ are quark flavor indices.
Constants $C_i^{\alpha\beta\gamma}$ weight the various Dirac components as required  to form an
irrep of the octahedral group.
When each quark field is projected to the space of the $N_{\rm eig}$ lowest eigenvalues of the covariant Laplacian
as in Eq.~(\ref{eq:project}),
 the correlation function becomes (with sums over repeated indices understood) 
 \begin{eqnarray}
 C_{ij}(t,t') =  \Phi_{i, k \ell m}^{\alpha\beta\gamma}(t) \,\,\,\Big\langle \widetilde{q}_{\alpha}^{(k)}(t)\widetilde{q}_{\beta}^{(\ell)}(t)  \widetilde{q}_{\gamma}^{(m)}(t)
\nonumber \\
\overline{\widetilde{q}}_{\bar{\alpha}} ^{(\bar{k})}(t')  \overline{\widetilde{q}}_{\bar{\beta}} ^{(\bar{\ell})}(t') 
\overline{\widetilde{q}}_{\bar{\gamma}} ^{(\bar{m})}(t') \Big\rangle \,\,\, \Phi_{j, \bar{k}\bar{\ell}\bar{m}}^{\bar{\alpha}\bar{\beta}\bar{\gamma} \dag} (t')
\label{eq:Cij}
\end{eqnarray}
where
\begin{equation}
   \Phi_{i, k \ell m}^{\alpha\beta\gamma}(t) = C_i^{\alpha\beta\gamma} \sum_{\bf x} \epsilon^{abc}  v_{a,\bf x}^{(k)}(t) v_{b,\bf x}^{(\ell)}(t)  v_{c,\bf x}^{(m)}(t). 
   \end{equation}
   
   Further reduction of the correlation functions involves contractions of  
   smeared quark fields that have the same quark flavor.  That is different for each  
   baryon but involves the same set of contractions as for unsmeared fields.  With the 
   smearing used here, each nonvanishing contraction yields a ``perambulator'' in the terminology of 
   Ref.~\cite{Peardon:2009}, i.e., 
   \begin{eqnarray}
  \tau ^{k \bar{k}}_{\alpha \bar{\alpha}} (t,t') &=&  \langle  \widetilde{q}_{\alpha}^{(k)}(t) \overline{\widetilde{q}}_{\bar{\alpha}} ^{(\bar{k})}(t') \rangle_A \nonumber \\
 &=& v_{b,\bf y}^{(k)\dag}(t)  \big(M^{-1} \big)^{bc}_{{\bf y z}, \alpha \bar{\alpha}} (t,t')  v_{c,\bf z}^{(\bar{k})}(t'),
     \end{eqnarray}
  where $M$ is the Dirac matrix and $\langle \cdots \rangle_A$ denotes evaluation with a single gauge configuration. 
    The perambulators are matrices in $N_{\rm eig} \times N_{\sigma}$ dimensions for each pair of source and sink times, t' and t.  As shown in Ref.~\cite{Peardon:2009}, $N_{\rm eig}=32$ provides smearing comparable to that
    based on Gaussian smearing  for $16^3\times 128$ lattices.  We use $N_{\rm eig}=32$ in this work.   The perambulator matrices provide the quark propagation from all 
    eigenvectors of the Laplacian at the source time to all eigenvectors at the 
sink time without reference to the operators that are used. 
    
    Matrices of correlation functions $C_{ij}(t,t')$ are obtained for four time slices $t'$.  We translate each of these correlation functions to $t' = 0$, yielding $C_{ij}(t,0)$, and average them for each configuration.
   The average over gauge configurations of  Eq.~(\ref{eq:Cij}) is calculated using the baryon operators
    $ \Phi_{i, k \ell m}^{\alpha\beta\gamma}(t)$ and $\Phi_{j, \bar{k}\bar{\ell}\bar{m}}^{\bar{\alpha}\bar{\beta}\bar{\gamma} \dag} (0)$ together with perambulators  $\tau ^{k \bar{k}}_{\alpha \bar{\alpha}} (t,0)$ for the relevant contractions.

    The variational method\cite{Michael:1985, Luscher:1990} is used to extract the energy levels.  We have exploited the fact that
    the matrices of correlation functions are real-valued within noise after removing a time-independent phase from each operator $B_i({\bf x},t)$ and the corresponding complex-conjugate phase from 
    $\overline{B}_j({\bf y},t)$.  Imaginary parts of the matrix elements are therefore dropped
     and that helps to reduce the overall noise.  The hermitian matrices $C_{ij}(t,0)$ become 
     real-symmetric matrices.  
 
    The generalized eigenvalue problem is solved at time $t^*$ to 
obtain eigenvectors of the correlation matrices.   Because matrix elements of the Hamiltonian involve
an average over configurations, we use the correlator matrix averaged
over configurations to obtain the eigenvectors as follows,
    \begin{equation}
    C_{ij}(t^*,0) u_j^{(n)}(t^*) = \lambda_n(t^*) C_{ij}(t_0,0) u_j^{(n)}(t^*),
    \label{eq:fixed_evecs}
    \end{equation}
    where $t_0$ is the normalization time for which the eigenvalues obey $\lambda_n(t_0) = 1$.
    Matrix indices $i$ and $j$ each take $N_{\rm op}$ values where $N_{\rm op}$ is the number of operators used.   
    For each gauge configuration, we then calculate matrix elements of $C_{ij}(t,0)$ in the basis of 
    fixed eigenvectors determined at time $t^*$, defining the ensemble of effective eigenvalues as follows,
    \begin{eqnarray}
      \widetilde{\lambda}_n(t)  = u_i^{(n)\dag }(t^*) C_{ij}(t,0)  u_j^{(n)}(t^*) .
      \label{eq:lambda_tilde}
    \end{eqnarray}
    At time $t=t^*$, the average over configurations of $\widetilde{\lambda}_n(t)$ is equal to $\lambda_n(t)$ because the average correlator matrix was used in Eq.~(\ref{eq:fixed_evecs}). At time $t=t_0$ 
    both the average $\widetilde{\lambda}_n(t)$ and $\lambda_n(t)$ equal one for the same reason.  
     The use of fixed eigenvectors provides a smooth time dependence in the
    diagonal correlation functions,  $\widetilde{\lambda}_n(t)$, and is consistent with
    the fact that the eigenvectors of the Hamiltonian should be independent 
    of time.  Differences between the fixed and exact eigenvectors contribute at second order
    to the difference between $\widetilde{\lambda}_n(t)$ and $\lambda_n(t)$ owing to the variational 
    nature of the calculation.  The accuracy of the fixed eigenvalues analysis has been checked by comparing with the exact eigenvalues.  The two analyses agree within uncertainties.
    
        There are two other reasons for using fixed eigenvectors.  One is that the eigenvalue problem  
        generally becomes ill-conditioned at late times owing to the
        exponential decrease of $\lambda_n(t) \propto e^{-E_n(t-t_0)}$ as $t$ becomes large.  The higher energy states tend to zero fastest so
        the condition number of the $N_{\rm op}\times N_{\rm op}$ matrix increases exponentially 
$\propto e^{(E_{\rm high} -E_{\rm low})(t-t_0)}$, where $E_{\rm high}$ and $E_{\rm low}$ are the largest and smallest 
        energies obtained from $N_{\rm op}$ operators.   
        For the $N_{\rm op} =$ 7 to 11 operators used, the condition number 
        becomes very large by $t \approx 15$.  Backward propagating baryon states can become 
        significant also at large times. When that happens the smallest eigenvalue can pass through 
        zero and become negative. The use of eigenvectors fixed at a time $t^*$ significantly earlier that $t \approx 15$ avoids  
         the conditioning problem.
        
        The second reason for using fixed eigenvectors is that at early times
        the eigenvectors generally are contaminated by contributions from states above the energy range that is
         determined by $N_{\rm op}$ operators. By choosing time $t_0$ to be as large as possible, 
         one reduces the contamination from higher states in the spectrum, which are 
         suppressed by factors involving $e^{-E t_0}$ as shown in Ref.~\cite{Blossier:2009}.  
        We usually set $t^* = t_0 +1$ in order to diagonalize a matrix
        that is effectively $e^{-H} \approx 1 - H + \cdots$, with eigenvectors dominated by
        the Hamiltonian, $H$.  Moreover, a  long fitting interval $(t_i, t_f)$ is required for
        accurate determination of the energies.  We  
        balance the requirements by choosing
        $t_0$ to be large while keeping $t^*$ small enough  
        to allow a well-conditioned determination of $N_{\rm op}$ eigenvectors.  The effective eigenvalues
        $\widetilde{\lambda}_n(t) $ are then fit as described in the next section in order to extract energies.

         \section{Analysis of nucleon spectra at $m_{\pi} =$ 392(4) MeV}
    \label{sec:N-spectrum-392}
 
Spectra have been calculated for the isospin $\frac{1}{2}$ (N) states, isospin $\frac{3}{2}$
($\Delta$) states and strangeness $-3$ ($\Omega$) states using lattices with the three sets of quark masses shown in Table~\ref{tab:quarkmasses}.   In this section we present a detailed discussion of our
analysis for the $N^*$ states at the lightest pion mass to show the procedure we employ
and to illustrate the quality of the data; the procedure for other cases is similar.

Table~\ref{tab:Nmasses} shows the results of fitting each nucleon diagonal correlation function, 
$\widetilde{\lambda}_n(t)$, by a two-exponential function of time as follows,
\begin{equation}
\lambda_{fit}(t) = (1-A) e^{-E(t-t_0)} + A e^{-E^{\prime}(t-t_0)},
\label{eq:Cfit}
\end{equation}
where $E < E^{\prime}$ so that $E$ is the energy of interest at large time.  The second
exponential term serves to model the contaminations arising from higher-energy states
at early times. The fit window $(t_i, t_f)$ is chosen such that coefficient $A$ is small. 
The values of $t_i$ and $t_f$ used are given in Table~\ref{tab:Nmasses} as are the values of $t_0$ and diagonalization 
time $t^*$ .  Although the lowest five energies are shown, the number of operators used is 7 to 11 in each irrep as shown in Table~\ref{tab:pruned_ops}.   
 Fits are performed for a jackknife ensemble of diagonal correlation functions calculated as in Eq.~(\ref{eq:lambda_tilde}), producing a jackknife ensemble of fit energies whose mean and standard deviation 
are given in the table.    

\begin{table}
\caption{Fit parameters for nucleon states at $m_{\pi} = 392(4)$ MeV. \label{tab:Nmasses}}
\begin{tabular}{lr}
\begin{tabular}{cccclllc}
\hline
IR &$ (t_0,t^*)$ & $ (t_i,t_f)$  &  $ E$     &      $1-A$     &    $ E^{\prime}$   & $\frac{\chi^2}{dof}$  ~~ \\ 
$G_{1g} $ &  (7,8) &   (7,31)  & 0.2085(19)   & 0.868(26)   & 0.427(42) & 1.63  \\ 
$G_{1g} $ &  (7,8) &   (7,24)  & 0.3545(51)   & 0.625(31)   & 0.559(14) & 0.49  \\ 
$G_{1g} $ &  (7,8) &   (5,22)  & 0.3675(110)   & 0.757(63)   & 0.685(44) & 0.66  \\ 
$G_{1g} $ &  (7,8) &   (5,18)  & 0.3831(95)   & 0.715(44)   & 0.725(28) & 1.02  \\ 
$G_{1g} $ &  (7,8) &   (4,15)  & 0.4205(88)   & 0.820(35)   & 0.817(34) & 1.39  \\ 
$G_{1g} $ &  (7,8) &   (4,14)  & 0.5320(124)   & 0.886(32)   & 0.991(51) & 0.83  \\ 
IR &$ (t_0,t^*)$ & $ (t_i,t_f)$  &  $ E$     &      $1-A$     &    $ E^{\prime}$   & $\frac{\chi^2}{dof}$  ~~ \\ 
$G_{1u} $ &  (7,9) &   (6,25)  & 0.2957(30)   & 0.840(24)   & 0.576(28) & 0.58  \\ 
$G_{1u} $ &  (7,9) &   (5,25)  & 0.3177(41)   & 0.895(22)   & 0.672(40) & 1.77  \\ 
$G_{1u} $ &  (7,9) &   (4,14)  & 0.4317(164)   & 0.758(59)   & 0.808(39) & 0.68  \\ 
$G_{1u} $ &  (7,9) &   (4,15)  & 0.4593(382)   & 0.720(135)   & 0.821(72) & 1.41  \\ 
$G_{1u} $ &  (7,9) &   (4,17)  & 0.4605(50)   & 0.914(16)   & 0.917(36) & 1.09  \\ 
$G_{1u} $ &  (7,9) &   (4,15)  & 0.4720(99)   & 0.806(34)   & 0.883(30) & 0.70  \\ 
IR &$ (t_0,t^*)$ & $ (t_i,t_f)$  &  $ E$     &      $1-A$     &    $ E^{\prime}$   & $\frac{\chi^2}{dof}$  ~~ \\ 
$H_g$  &  (8,9) &   (7,25)  & 0.3541(54)   & 0.650(46)   & 0.547(17) & 0.89  \\ 
$H_g$  &  (8,9) &   (7,25)  & 0.3643(29)   & 0.840(12)   & 0.633(12) & 1.37  \\ 
$H_g$  &  (8,9) &   (7,20)  & 0.3735(93)   & 0.760(82)   & 0.610(52) & 0.62  \\ 
$H_g$  &  (8,9) &   (5,17)  & 0.4053(56)   & 0.874(29)   & 0.740(36) & 0.53  \\ 
$H_g$  &  (8,9) &   (5,17)  & 0.4092(64)   & 0.898(35)   & 0.753(54) & 0.37  \\ 
$H_g$  &  (8,9) &   (5,17)  & 0.4129(77)   & 0.886(35)   & 0.808(54) & 0.92  \\ 
IR &$ (t_0,t^*)$ & $ (t_i,t_f)$  &  $ E$     &      $1-A$     &    $ E^{\prime}$   & $\frac{\chi^2}{dof}$  ~~ \\ 
$H_u$  &  (8,9) &   (7,25)  & 0.3037(34)   & 0.738(38)   & 0.485(18) & 1.08  \\ 
$H_u$  &  (8,9) &   (7,25)  & 0.3065(44)   & 0.724(50)   & 0.481(21) & 0.72  \\ 
$H_u$  &  (8,9) &   (7,24)  & 0.3203(25)   & 0.832(22)   & 0.578(26) & 0.84  \\ 
$H_u$  &  (8,9) &   (7,24)  & 0.3383(61)   & 0.800(61)   & 0.589(56) & 1.04  \\ 
$H_u$  &  (8,9) &   (4,18)  & 0.4516(102)   & 0.869(31)   & 0.855(34) & 0.63  \\ 
$H_u$  &  (8,9) &   (4,18)  & 0.4628(81)   & 0.919(24)   & 0.900(46) & 1.35  \\ 
IR &$ (t_0,t^*)$ & $ (t_i,t_f)$  &  $ E$     &      $1-A$     &    $ E^{\prime}$   & $\frac{\chi^2}{dof}$  ~~ \\ 
$G_{2g} $ &  (7,8) &   (5,22)  & 0.3870(69)   & 0.814(43)   & 0.737(45) & 1.71  \\ 
$G_{2g} $ &  (7,8) &   (5,18)  & 0.3930(85)   & 0.711(53)   & 0.675(29) & 1.64  \\ 
$G_{2g} $ &  (7,8) &   (5,18)  & 0.4006(80)   & 0.725(47)   & 0.704(29) & 2.67  \\ 
$G_{2g} $ &  (7,8) &   (4,18)  & 0.4278(73)   & 0.832(30)   & 0.852(33) & 1.12  \\ 
$G_{2g} $ &  (7,8) &   (4,14)  & 0.5405(205)   & 0.887(69)   & 1.000(112) & 1.07  \\ 
$G_{2g} $ &  (7,8) &   (4,14)  & 0.5701(129)   & 0.882(34)   & 1.032(49) & 0.82  \\ 
IR &$ (t_0,t^*)$ & $ (t_i,t_f)$  &  $ E$     &      $1-A$     &    $ E^{\prime}$   & $\frac{\chi^2}{dof}$  ~~ \\ 
$G_{2u} $ &  (6,8) &   (6,23)  & 0.3407(45)   & 0.746(42)   & 0.607(38) & 0.70  \\ 
$G_{2u} $ &  (6,8) &   (4,15)  & 0.4586(76)   & 0.825(28)   & 0.944(38) & 2.41  \\ 
$G_{2u} $ &  (6,8) &   (4,16)  & 0.4802(53)   & 0.875(23)   & 0.949(44) & 0.69  \\ 
$G_{2u} $ &  (6,8) &   (5,16)  & 0.4958(74)   & 0.841(26)   & 0.930(37) & 1.60  \\ 
$G_{2u} $ &  (6,8) &   (5,16)  & 0.4992(63)   & 0.865(19)   & 1.092(52) & 1.23  \\ 
$G_{2u} $ &  (6,8) &   (5,15)  & 0.5239(151)   & 0.917(56)   & 1.367(358) & 1.96  \\ 
\hline
\end{tabular}
\end{tabular}
\end{table}

\begin{figure*}[h]
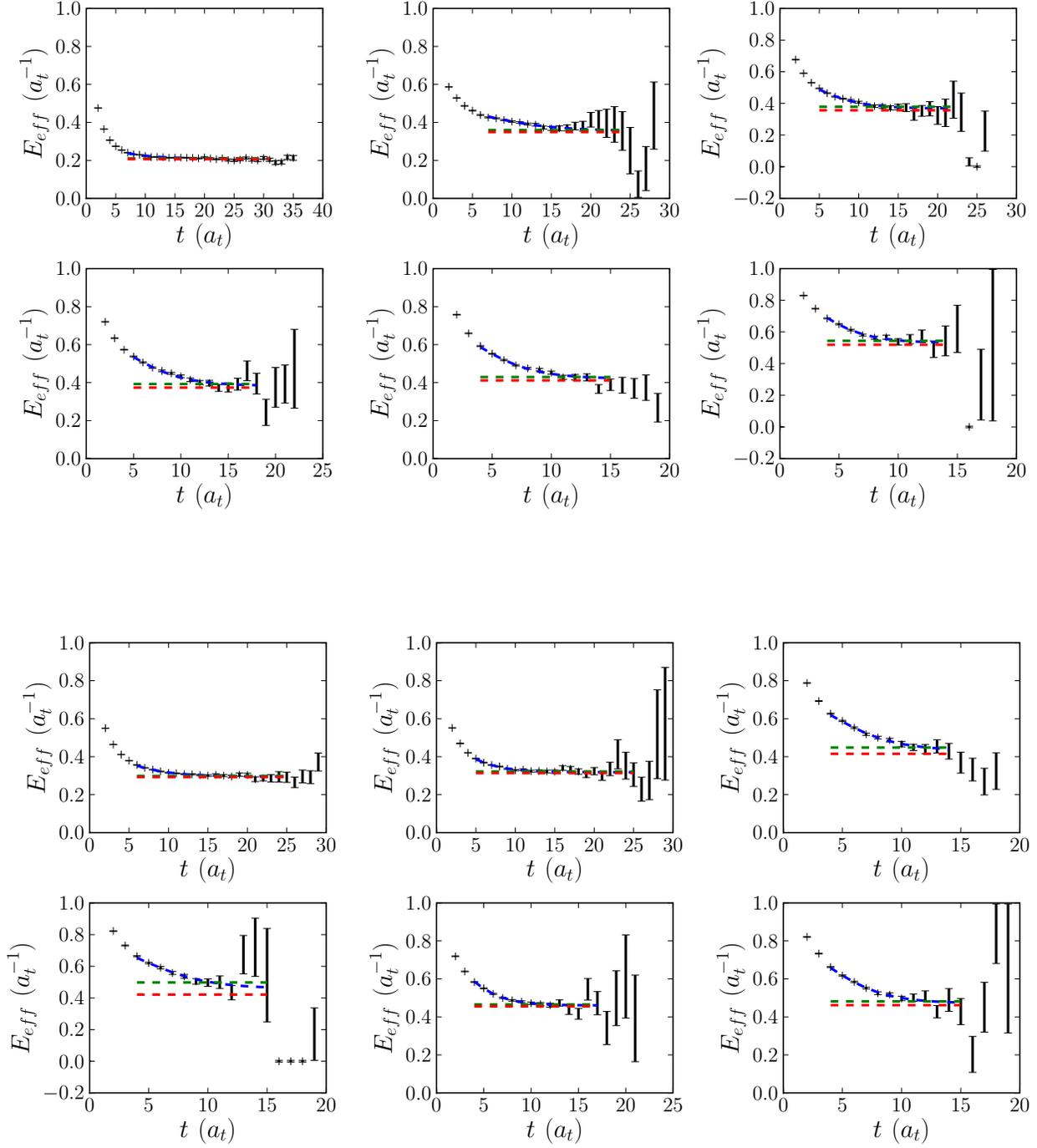

\ig[bb= 150 100 500 750]{nucleon_m0840_meffplots_G1.pdf}

\vspace{-1in}

\caption{ Nucleon $G_{1g}$ effective energies are shown for the lowest states in the upper 
six graphs.  The effective energy increases from left to right along the first row and continues to increase from left to right along the second row.  The lower six graphs show 
 nucleon $G_{1u}$ effective energies increasing in the same pattern.
Calculations are for $m_{\pi}= 392(4)$ MeV. Vertical bars show the effective energy and the curved dashed line shows the effective energy calculated from the fit function.  Horizontal dashed lines show the
fit results for $E \pm\sigma$ and their extent shows the fitting interval $(t_i,t_f)$.\label{fig:meffplots_G1}}
\end{figure*}

Plots of the nucleon effective energies, calculated as 
\begin{equation}
E_{\rm eff}(t) = \frac{1}{2} ln\left( \frac{\widetilde{\lambda}(t-1)}{\widetilde{\lambda}(t+1)}\right),
\label{eq:Eeff}
\end{equation}
are shown in Figure~\ref{fig:meffplots_G1} for the $G_{1g}$ and $G_{1u}$ irreps, Fig.~\ref{fig:meffplots_H} for the $H_g$ and $H_u$ irreps, and Fig.~\ref{fig:meffplots_G2} for the
$G_{2g}$ and $G_{2u}$ irreps.  These plots show the values of $E_{\rm eff}$ obtained from 
Eq.~(\ref{eq:Eeff}) as vertical bars and $E_{\rm eff}$ calculated using the fit function of
Eq.~(\ref{eq:Cfit}) in place of $\widetilde{\lambda}(t)$ in Eq.~(\ref{eq:Eeff}) as dashed lines.  Comparison of the dashed lines with the bars from the lattice ensembles
shows the usefulness of two-exponential fits.  The term $Ae^{-E^{\prime}(t-t_0)}$ 
models the contributions of higher energy states at early times allowing the exponential term $(1-A)e^{-E(t-t_0)}$ to be determined over a larger fit window $(t_i, t_f)$ than would be possible using a single exponential.  
 Fit energy $E$ and uncertainty of the fit energy, $\sigma$, are shown by dashed horizontal lines
 at $E +\sigma$ and $E-\sigma$ extending over the fit window. Note that the fits over a long time interval provide smaller uncertainties when compared with the variations of the effective masses.  The latter have contributions from higher states at early times and local fluctuations at late times because they are calculated from the correlation function at next-to-nearest times. 
Note also that the statistics allow credible determinations of six energy levels in each irrep.

\begin{figure*}[h]
\ig[bb= 150 150 500 750]{nucleon_m0840_meffplots_H.pdf}

\vspace{-1in}

\caption{ Nucleon $H_{g}$ effective energies are shown for the lowest states in the upper 
six graphs and nucleon $H_{u}$ effective energies are shown in the lower six graphs 
with effective energies increasing in the same pattern as in Fig.~\ref{fig:meffplots_G1}.
Calculations are for $m_{\pi} = 392(4)$ MeV. Vertical bars show the effective energy and the curved dashed line shows the effective energy calculated from the fit function. Horizontal dashed lines show the
fit results for $E \pm\sigma$ and their extent shows the fitting interval $(t_i,t_f)$. 
\label{fig:meffplots_H}}
\end{figure*}
\begin{figure*}[h]
\ig[bb= 150 150 500 750]{nucleon_m0840_meffplots_G2.pdf}
\vspace{-1in}

\caption{ Nucleon $G_{2g}$ effective energies are shown for the lowest states in the upper 
six graphs and nucleon $G_{2u}$ effective energies are shown in the lower six graphs
with effective energies increasing in the same pattern as in Fig.~\ref{fig:meffplots_G1}.
Calculations are for $m_{\pi}= 392(4)$ MeV.  Vertical bars show the effective energy and the curved dashed line shows the effective energy calculated from the fit function. Horizontal dashed lines show the
fit results for $E \pm\sigma$ and their extent shows the fitting interval $(t_i,t_f)$. 
\label{fig:meffplots_G2}}
\end{figure*}

The same process has been used to obtain $N$, $\Delta$ and $\Omega$ energies 
at three values of $m_{\pi}$.  The results are given in summary form in the next section.

\section{$N$, $\Delta$ and $\Omega$ spectra at three pion masses}
\label{sec:N-D-Om-spectra}

     \begin{table}
\caption{ Multiparticle thresholds on the lattice for isospin $I$ and strangeness $S$ are shown for each value of $m_{\pi}$ in MeV.  The thresholds are based on the sum of energies of the
particles with no interactions. \label{tab:thresholds} }
   \begin{center}
     \begin{tabular}{|c|c|c|c|ccc|}
      \hline
   I & S & IR              &      State           &  $m_{\pi}$=392  &  $m_{\pi}$=438 &  $m_{\pi}$=521 \\  
   \hline
  $\frac{1}{2}$, $\frac{3}{2}$ & 0 & $G_{1g}$ & $(N\pi\pi)_{s-wave} $  &  1965(19) &  2107(9) &  2352(9) \\
     $\frac{1}{2}$, $\frac{3}{2}$ & 0 &$H_g$      & $(N\pi)_{p-wave}$      &  2089(18) &  2133(7) &  2220(6) \\
    $\frac{1}{2}$, $\frac{3}{2}$ & 0 & $G_{2g}$ & $(\Delta \pi)_{p-wave}$& 2361(21)&  2375(9) &  2446(9) \\
    $\frac{1}{2}$, $\frac{3}{2}$ & 0 & $G_{1u}$ & $(N\pi)_{s-wave}$        &  1573(16) &  1669(7) & 1831(6) \\
     $\frac{1}{2}$, $\frac{3}{2}$ & 0 &$H_u$      & $(\Delta \pi)_{s-wave}$&  1875(18) &  1934(9) &  2075(9) \\
    $\frac{1}{2}$, $\frac{3}{2}$ & 0 & $G_{2u}$ &  $(N \pi)_{d-wave} $    &  2089(18) &  2133(7) &  2220(6) \\
\hline
   0 & -3 &  $G_{1g}$ & $(\Xi K)_{p-wave} $  &  2337(20) &	2351(11)	 &  2348(15)\\
   0 & -3 &   $H_g$      & $(\Omega\pi\pi)_{s-wave}$      &  2456(20)	 & 2549(9) & 2714(9)  \\
   0 & -3 &   $G_{2g}$ & $(\Xi K)_{p-wave}$& 2337(20) &	2351(11)	 &  2348(15) \\
    0 & -3 &  $G_{1u}$ & $(\Xi K)_{s-wave}$   & 1904(18)	 & 1949(12) & 1991(17)     \\
    0 & -3 &  $H_u$      & $(\Omega \pi)_{s-wave}$&  2064(16)	 & 2111(6) & 2193(6) \\
    0 & -3 &  $G_{2u}$ &  $(\Xi K)_{d-wave} $    & 2337(20) &	2351(11) & 2348(15) \\
\hline
\end{tabular}   
 \end{center}   \end{table}

The goal of determining the spectra of baryon resonances from lattice QCD
requires an increasing elaborate analysis as the limit of physical pion mass and large volume is approached.  Although all lattice states have discrete energies at any finite volume, the energies
correspond to single-particle states, interacting multi-particle states and mixtures thereof.
At a minimum, one needs to resolve all the states
up to some energy and identify them as predominantly resonances or predominantly scattering states.   
The repulsion or attraction of multi-particle energy levels at finite volume can be related to the 
momentum-dependent phase shifts; the (model-dependent) resonance parameters can then be extracted through, say, a Breit-Wigner fit to the phase shift.  It also should be noted that experimental
resonances generally involve mixtures of single-particle states and multi-particle states and in some cases there may be a linear combination of multi-particle states that produces features similar 
to those of a resonance.

 Although we have spectra for three values of $m_{\pi}$, we cannot clearly delineate multi-particle
 states in the spectrum and are unable to obtain the energy-dependent phase shift;  that analysis must await the introduction of a broader basis of operators.  In the following, we do not attempt to perform 
 a chiral extrapolation on the spectrum.  The couplings of the excited states are in general unknown, and we are
 performing calculations in a region in which, as we will see below, multi-particle contributions are expected. 

\subsection{Nucleon spectra}

Spectra for isospin $\frac{1}{2}$ states (N states) are summarized in Figure~\ref{fig:N-boxplots} for each lattice
irrep and for $m_{\pi} =$ 392, 438 and 521 MeV.   We also show the isospin $\frac{1}{2}$ two-star, three-star and four-star experimental resonances
with $J^P$ values that have a subduction to the lattice irrep.  Experimental resonances~\cite{PDG:2008} are shown by boxes 
  in columns labeled by their $J^P$ values with the height of the box equal to the full decay width of the resonance.  An inner box (color aqua) shows the uncertainty in the Breit-Wigner resonance 
  energy.  The lattice results are shown as colored boxes 
 with height equal to 2$\sigma$ in the columns labeled by values of $m_{\pi}$, where $\sigma$ is 
 the statistical uncertainty of the fit energy. The lattice energies have been converted to
 MeV units by the formula,	
 \begin{equation}
            E = 1672.45 \Bigr( \frac{E a_t}{m_{\Omega} a_t}\Bigr)^{\rm latt}. 
            \label{eq:scale}
  \end{equation} 
 Here the ratio of a lattice energy and the lattice $m_{\Omega}$ is calculated for each value
 of $m_{\pi}$ and then is scaled by the empirical mass of the $\Omega$-baryon, 1672.45 MeV.
 Triangles to the right of the lattice spectra for each value
 of $m_{\pi}$ in Fig.~\ref{fig:N-boxplots} point to the threshold for scattering states (multi-particle states) 
 at that value of $m_{\pi}$.  Most of the excited states have energies higher than the thresholds for
 scattering states.

 As an example, we discuss the $G_{1g}$ plot, shown in the upper left hand panel of 
Figure~\ref{fig:N-boxplots}.  The experimental spectrum  contains three low-lying $\frac{1}{2}^+$ states 
together with one $\frac{7}{2}^+$ state.  For the lattice calculation, we show the six lowest
energy states at each pion mass.  Each experimental state
shown has a subduction to isospin $\frac{1}{2}$, $G_{1g}$, and should occur in the lattice QCD
spectrum for each value of $m_{\pi}$.  The lattice states should correspond 
 to subductions of continuum states.  The relevant 
 continuum states consist of the experimental resonances and scattering states with $J^P$ values
 that have subductions to the lattice irrep.  Thresholds for scattering states are shown in Fig~\ref{fig:N-boxplots}  and are listed in 
 Table~\ref{tab:thresholds}.  The listed multiparticle states 
are assigned to lattice irreps following Ref.~\cite{Moore:2006}.
 Because the spatial lattice is a cube measuring about 1.96 fm on a side, 
 the momentum is restricted to discrete values with the smallest nonzero one being 630 MeV.  Consequently, the scattering states with nonzero momenta occur at higher energies. 

The general pattern seen in Figure~\ref{fig:N-boxplots} is that lattice states have high energies that decrease  toward the experimental resonances as $m_{\pi}$ is decreased.   The density of lattice states increases with increasing energy, but we restrict the analysis to the lowest six states.
 That we are able to extract six energies in each lattice irrep is testament to the effectiveness of the
  smearing procedure based on eigenvectors of the lattice Laplacian.
  Note, however, that we cannot distinguish
  clearly between single- and multi-particle contributions to the spectrum in this calculation. 
  
The nucleon ground state shows up as the lowest state in the $G_{1g}$ lattice irrep.  Its energy decreases in a regular manner toward the experimental value shown
in the $\frac{1}{2}^+$ column as $m_{\pi}$ decreases.    
     The first excited state decreases toward the Roper resonance, $N^*(1440, \frac{1}{2}^+)$ but remains 
     above about 1900 MeV for the pion masses used in this work.   
       
       In the negative-parity $G_{1u}$ spectra, there are two low-lying states at each value of $m_{\pi}$
       and they tend toward the experimental resonances $N(1535, \frac{1}{2}^-)$ and $N(1650, \frac{1}{2}^-)$ as $m_{\pi}$ decreases.  A number of higher states also tend toward the energy of the $N(2200, \frac{7}{2}^-)$ resonance
       or to scattering states such $(N\pi)_{s-wave}$, $(N\pi)_{p-wave}$ and so on.   
       
   Before we discuss the $H$ and $G_2$ states, it is worth noting that isolated $G_2$ states do not 
   correspond to any physical state.  Because $G_2$ has minimum spin $\frac{5}{2}$, there must be at least six 
   linearly independent components in the continuum limit.  Each $G_2$ state must have a partner $H$ state with the same parity in order to have an interpretation as a physical state.  However, on the lattice discretization effects can cause the $H$ and $G_2$ partner states to have different energies at ${\cal O}(a^2)$.     
       
       In the $H_g$ spectra there are five experimental resonances: $N(1720, \frac{3}{2}^+)$ and $N(1900, \frac{3}{2}^+)$, $N(1680, \frac{5}{2}^+)$, $N(2000, \frac{5}{2}^+)$ and $N(1990, \frac{7}{2}^+)$.  The lattice states tend as a group toward these energies as $m_{\pi}$ decreases. 
       
       In the $H_u$ spectra there are four low-lying lattice states near 1800 MeV.
          The threshold for scattering states is near the same energy as this group of 
      lattice states.  Three low-lying experimental resonances are present:  N(1530, $\frac{3}{2}^-$),
N(1650, $\frac{3}{2}^-$) and N(1675, $\frac{5}{2}^-$).
           
 In our $N_f$=2 analysis of Ref.~\cite{Bulava:2009}, we obtained three low-lying $H_u$ states with larger uncertainties.  Otherwise the
 low-lying lattice states agree reasonably well.  A test of stability was performed by 
omitting $H_u$ operators to obtain sets of $N_{\rm op}=$ 6, 7, 8 and 9.  Spectra were 
calculated for each of these and the results show three $H_u$ states 
near 2000 MeV when we use 6 or 7 operators and four $H_u$ states when we use 8 or 
9 operators.  This behavior suggests that one state is only resolved with the larger number of operators.  The operators that are responsible for the appearance of the
fourth state are of the triply-displaced-T type.   
We also have studied the stability of the spectrum at $m_{\pi}$ = 561 MeV by varying $t_0$, 
keeping $t^* = t_0+1$ and using all operators.  For $t_0=$ 2, 3, 4, 5, 6, 7 and 8 we observe four low-lying states in the nucleon $H_u$ spectrum.
The presence of a fourth state is robust when all operators are used.  
  
  Four low-lying $H_u$ states are consistent with experiment if  one is a scattering state.  We have not previously found any evidence for a scattering state with our three-quark operators, but they should be present.  Work that is in progress aims to 
identify the scattering states by using operators designed to couple to them directly.   
      
       The pattern that is used to identify a spin $\frac{5}{2}^-$ state on the lattice is a pair of states in the $G_{2u}$ and $H_u$ irreps that
      become degenerate in the continuum limit.  A candidate for this pattern is present in our $H_u$ and $G_{2u}$ spectra: the lowest $G_{2u}$ state and one of the four $H_u$ states at essentially the
      same energy.   The pattern for spin-$\frac{7}{2}$ is a triplet of $G_1$, $H$ and $G_2$ states at essentially the same energy.  There are candidates for this pattern in the positive-parity spectra.  However, the presence of scattering states makes a secure identification difficult.  

\begin{figure*}[b]
\begin{tabular}{ccc}
\ig[width=1.0\tw] {nucleon_boxplots_all.pdf}
\end{tabular}
\vspace{-0.3in}\caption{Spectra for isospin $\frac{1}{2}$ (nucleon family) at three values of $m_{\pi}$ are compared with experimental spectra.   Plots in the first row show $G_{1g}$ and $G_{1u}$ lattice irreps, plots in the
second row show $H_g$ and $H_u$ irreps and plots in the third row show $G_{2g}$ and $G_{2u}$ irreps.  Columns labeled by $m_{\pi} = $ 392, 438 and 521  
show lattice spectra at those values of $m_{\pi}$.  Two, three and four-star experimental resonances 
are shown in columns labeled by their $J^P$ values.  Each $J^P$ value listed has a subduction to the lattice irrep shown.  
Each box for an experimental resonance has height equal to the full decay width and an inner box (color aqua) showing the uncertainty in the Breit-Wigner energy.  Triangles to the right of lattice spectra point to
the threshold for scattering states at that value of $m_{\pi}$. 
\label{fig:N-boxplots}} 
\end{figure*}
%

\begin{figure*}[b]
\begin{tabular}{ccc}
\ig[width=1.0\tw] {delta_boxplots_all.pdf}
\end{tabular}
\vspace{-0.3in}\caption{Spectra for isospin $\frac{3}{2}$ ($\Delta$ family) at three values of $m_{\pi}$ are compared with experimental spectra.   Plots in the first row show $G_{1g}$ and $G_{1u}$ lattice irreps, plots in the
second row show $H_g$ and $H_u$ irreps and plots in the third row show $G_{2g}$ and $G_{2u}$ irreps.  Columns labeled by $m_{\pi} = $ 392, 438 and 521 
show lattice spectra at those values of $m_{\pi}$.  Two, three and four-star experimental resonances 
are shown in columns labeled by their $J^P$ values.  Each $J^P$ value listed has a subduction to the lattice irrep shown.  
Each box for an experimental resonance has height equal to the full decay width and an inner box (color aqua) showing the uncertainty in the Breit-Wigner energy.    Triangles to the right of lattice spectra point to
the threshold for scattering states at that value of $m_{\pi}$.
    \label{fig:D-boxplots}} 
\end{figure*}
\begin{figure*}[b]
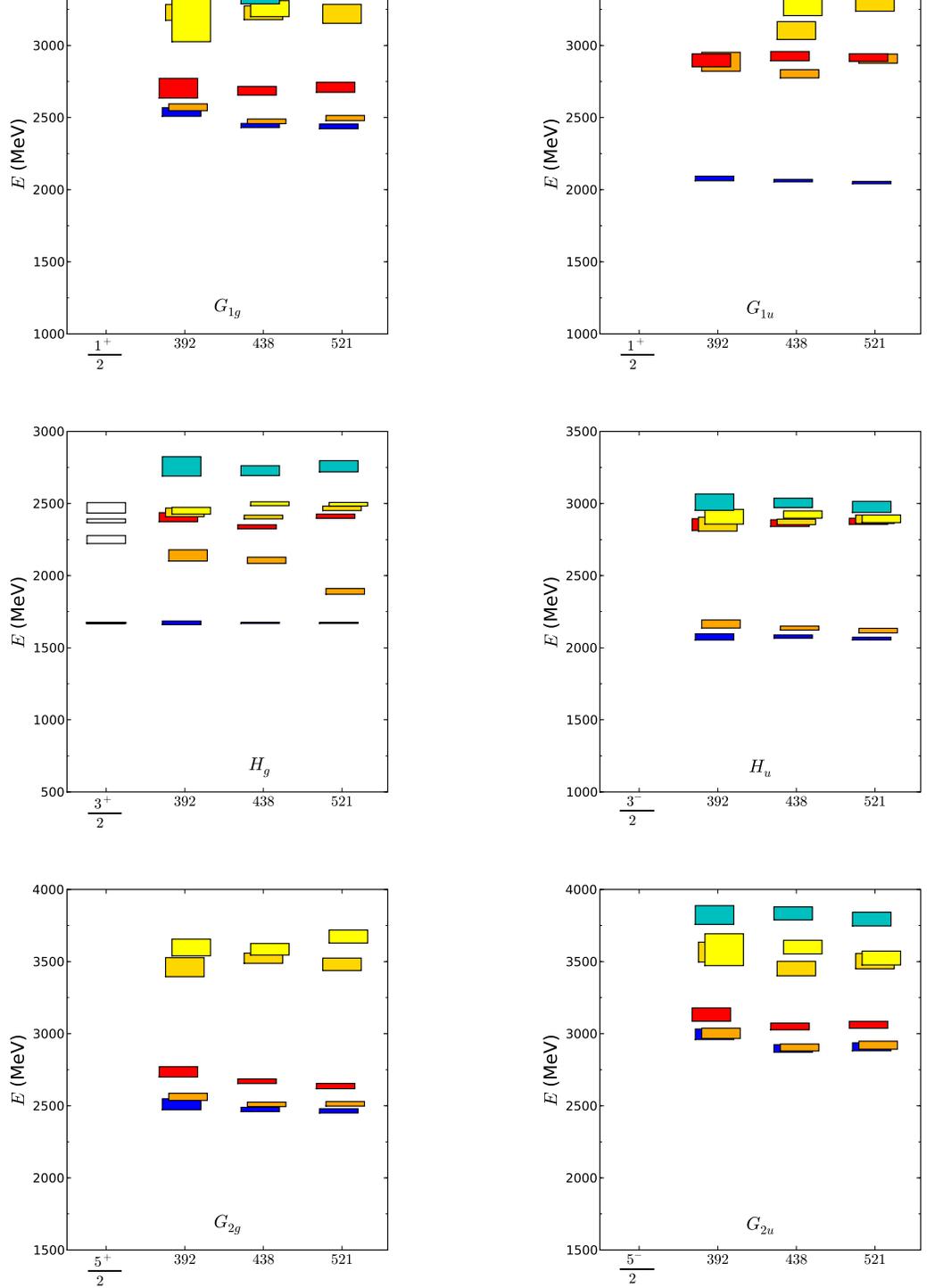

\begin{tabular}{cc}
\ig[width=1.0\tw] {omega_boxplots_all.pdf}
\end{tabular}
\vspace{-0.3in}\caption{Spectra for isospin 0, strangeness -3 ($\Omega$ family) at three values of $m_{\pi}$ are compared with experimental spectra.   Plots in the first row show $G_{1g}$ and $G_{1u}$ irreps, plots in the
second row show $H_g$ and $H_u$ irreps and plots in the third row show $G_{2g}$ and $G_{2u}$ lattice irreps.  Columns labeled by $m_{\pi} = $ 392, 438 and 521 
show lattice spectra at those values of $m_{\pi}$.  Two, three and four-star experimental resonances 
are shown in columns labeled by their $J^P$ values.  Each $J^P$ value listed has a subduction to the lattice irrep shown.  For the $\Omega$, the spins and parities of the experimentally observed states 
other than the lightest are not clearly determined; for comparison, we assign the states to $\frac{3}{2}^+$,
with the heights of the boxes indicating the widths.   Triangles to the right of lattice spectra point to
the threshold for scattering states at that value of $m_{\pi}$.
  \label{fig:Om-boxplots}} \vspace*{0.2in}
\end{figure*}

\subsection{$\Delta$ spectra}
 
The $\Delta$ spectra are shown in Fig.~\ref{fig:D-boxplots}.  The general features are the same as for the nucleon spectra: the lattice states are high and they tend toward the 
experimental resonances as $m_{\pi}$ decreases.  The $\Delta(1232, \frac{3}{2}^+)$ ground state appears as the lowest state in the $H_g$ spectra.  
  The next higher $H_g$ state is close to the ground state but appears to tend toward the $\Delta(1600,\frac{3}{2}^+)$ state in the experimental spectrum.  
  
  The lowest two $G_{1g}$ states near 2200 MeV appear to be somewhat high but consistent with the experimental resonances
  $\Delta(1910, \frac{1}{2}^+)$ and $\Delta(1950, \frac{7}{2}^+)$.   Candidates for  spin-$\frac{7}{2}^+$ partner states are
  present in the 2200 MeV to 2400 MeV range in $H_g$ and $G_{2g}$ spectra but the pattern is a weak
  match for the expected degeneracy in the continuum limit.  Possibly the small volume used is causing large splittings. 
  
 One of the two lowest $H_u$ states corresponds reasonably well to the $\Delta(1700,\frac{3}{2}^-)$ resonance. The other one should correspond to the $\Delta(1930,\frac{5}{2}^- )$ resonance.  However, a suitable partner state for spin $\frac{5}{2}^-$ is not seen in the $G_{2u}$ spectrum: the lowest such state is near 2600 MeV.   A similar result is found in the $G_{2g}$ spectrum with the lowest state being close to 2300 MeV, well above the energy of the $\Delta(1905, \frac{5}{2}^+)$ resonance. This suggests that the volume may be small,
 particularly for the 
$G_2$ lattice states.  In quark models~\cite{Capstick:2000qj, Isgur:1978wd},
excited states typically have larger radii. Our lattice
is about 1.8 fm in extent and a state with a radius of more than 1fm is not expected to be determined well.

\subsection{$\Omega$ spectra}

The spectra for excited $\Omega$ states are shown in Fig.~\ref{fig:Om-boxplots}.  The $\Omega(1672, 
    \frac{3}{2}^+)$  ground state has been used to set the scale
for baryon masses so is reproduced perfectly.   Experimental resonances above the ground state do not have spin-parity assignments.  In the quality rating of resonances of  Ref.~\cite{PDG:2008}, $\Omega(2250)$ is rated as a three-star resonance while $\Omega(2380)$ and $\Omega(2470)$ are rated as two-star resonances. The strange-quark mass is at its physical value
in our calculations and the dependence on the pion mass is expected to be smaller than for other
resonances. Consistent with this the overall pattern of excited states varies little
with $m_{\pi}$.   A noteworthy exception is the first $H_g$ excited state, whose energy increases from
about 1800 MeV to 2100 MeV as the pion mass decreases.   

We have considered whether the lattice $\Omega$ spectra can provide a useful guide for assignment
of spins and parities.  We find 11 strangeness -3 states with energies near or below 2500 MeV.
Some of those states may be candidates for scattering states rather than resonances.  Thresholds
for scattering states are shown for each value of $m_{\pi}$ in Figure~\ref{fig:Om-boxplots}.   However we cannot  determine whether or not our spectra contain scattering states. 

A reasonably good agreement between the lattice and experimental spectra is obtained if the
first excited experimental resonance is assigned to $\frac{3}{2}^+$.  Beyond that there are several possibilities.  In Fig.~\ref{fig:Om-boxplots} all of the experimental resonances have been
shown in the $\frac{3}{2}^+$ column that appears in the plot of $H_g$ spectra.  However, a convincing assignment is not 
possible because many features of the spectra are not explained.
This issue will be revisited when good operators for scattering states are available. 
 
    \section{Summary}  \label{sec:summary}
    
   This work represents a milestone in our long-term research program aimed at
   determining the spectra of baryons in QCD.  It provides the first spectrum for 
   N, $\Delta$ and $\Omega$ baryons based on $N_f=2+1$ QCD with high
   statistics.   A large number of baryon operators is used to calculate
   matrices of correlation functions.  They are analyzed using the variational method
   with fixed eigenvectors.  
   The analysis provides three spectra at pion masses, 
   $m_{\pi}$ =  392(4) MeV, 438(3) MeV and 521(3) MeV. 
       
   The lattice volume and pion masses used give considerably higher
   energies than the experimental resonance energies.  However, there is reasonable agreement of the overall 
   pattern of lattice and experimental states.  One exception is that almost all $G_2$ states 
   are much too high.  That may be caused by a volume that is too small for
   highly excited states.
       
   We find candidates for scattering states that have not shown up in our previous
   analyses based on $N_f=2$ QCD or quenched QCD.  We also find more excited state contamination
   in the effective-mass plots than was the case for quenched QCD.  We expect that when appropriate operators are
   used to identify multiparticle states, the spectra will be cleaner.  
   
   The spectrum of excited states of the $\Omega$ baryons has been calculated for the first time.
   We do not find a close enough agreement between lattice and experimental
   excited states to allow a convincing assignment of the unknown spins and parities.
   
   Our main conclusion is that the program to determine baryon spectra from lattice QCD is 
 expected to produce reasonable explanations of the nucleon, $\Delta$ and $\Omega$ spectra once calculations are extended to smaller pion masses, larger
 volumes and operators designed to couple to scattering states directly.
 Stochastic estimation of the quark propagators will allow use of the distillation method with larger volumes.\cite{Morningstar:2010}  
   
\acknowledgements This work was done using the Chroma software suite~\cite{Edwards:2005} on clusters at Jefferson Laboratory and the Fermi National Accelerator Laboratory using time awarded under the USQCD Initiative.
JB and CM acknowledge support from U.S. National Science Foundation Award 
PHY-0653315.  EE and SW acknowledge support from U.S. Department of Energy contract 
DE-FG02-93ER-40762.  HL acknowledges support from U.S. Department of Energy contract
DE-FG03-97ER4014.  BJ, RE and DR acknowledge support from U.S. Department of Energy contract DE-AC05-84ER40150, under which Jefferson Science Associates, LLC, manages and operates Jefferson Laboratory.  The U.S. Government retains a non-exclusive, paid-up, irrevocable,
world-wide license to publish or reproduce this manuscript for U.S. Government purposes.

\end{document}